\documentclass[aps,pra,reprint,showpacs,showkeys,showpacs,times,superscriptaddress]{revtex4-1}
\usepackage{amsmath,amssymb,amsthm,times,graphics,graphicx,bm,subfigure}

\usepackage{amssymb,amsmath}
\usepackage{graphicx}
\usepackage{color}
\usepackage{hyperref}
\usepackage{listings}
\hypersetup{
    bookmarks=true,         
    unicode=false,          
    pdftoolbar=true,        
    pdfmenubar=true,        
    pdffitwindow=false,     
    pdfstartview={FitH},    
    pdfauthor={Kevin Multani},     
    colorlinks=true,       
    linkcolor=blue,          
    citecolor=red,        
}
\graphicspath{{Figures/}}

\usepackage{pifont}

\newcommand{\ket}[1]{\vert#1\rangle}
\newcommand{\bra}[1]{\langle#1\vert}

\newcommand{\cp}[2]{\text{CPHASE}_{{#1},{#2}}}

\begin{document}

\definecolor{dkgreen}{rgb}{0,0.6,0}
\definecolor{gray}{rgb}{0.5,0.5,0.5}
\definecolor{mauve}{rgb}{0.58,0,0.82}

\lstset{frame=tb,
  	language=Matlab,
  	aboveskip=3mm,
  	belowskip=3mm,
  	showstringspaces=false,
  	columns=flexible,
  	basicstyle={\small\ttfamily},
  	numbers=none,
  	numberstyle=\tiny\color{gray},
 	keywordstyle=\color{blue},
	commentstyle=\color{dkgreen},
  	stringstyle=\color{mauve},
  	breaklines=true,
  	breakatwhitespace=true
  	tabsize=3
}

\title{Structure of multipartite entanglement in random cluster-like photonic systems}
\author{Mario A. Ciampini}
\affiliation{Dipartimento di Fisica, Sapienza Universit\`a di Roma, P.le Aldo Moro 5, 00185, Rome, Italy}
\author{Paolo Mataloni}
\affiliation{Dipartimento di Fisica, Sapienza Universit\`a di Roma, P.le Aldo Moro 5, 00185, Rome, Italy}
\author{Mauro Paternostro}
\affiliation{Centre for Theoretical Atomic, Molecular and Optical Physics, School of Mathematics and Physics, Queen's University Belfast, Belfast BT7 1NN, United Kingdom}
\date{\today}

\begin{abstract}
Quantum networks are natural scenarios for the communication of information among distributed parties, and the arena of promising schemes for distributed quantum computation. Measurement-based quantum computing is a prominent example of how quantum networking, embodied by the generation of a special class of multipartite states called {\it cluster states}, can be used to achieve a powerful paradigm for quantum information processing. Here we analyze randomly generated cluster states in order to address the emergence of multipartite correlations as a function of the density of edges in a given underlying graph. We find that the most widespread multipartite entanglement does not correspond to the highest amount of edges in the cluster. 
We extend the analysis to higher dimensions, finding similar results, which suggest the establishment of \emph{small world} structures in the entanglement sharing of randomised cluster states, which can be exploited in engineering more efficient quantum information carriers.   
\end{abstract}

\maketitle
\section{Introduction}
In 1929, the Hungarian author F. Karinthy famously set out the concept of {\it six degrees of separation}~\cite{6}, the conjecture according to which any two living entities on Earth are {\it distant} by no more than five intermediate steps. This concept was reprised and developed later on more rigorous sociological and statistical grounds. Remarkably, for instance, a variation of the {\it six degrees} was unveiled by the group of A.-L. Barabasi in 1999~\cite{barabba}, who predicted that any page in the World Wide Web can be reached from any other one with only nineteen intermediate steps (or clicks) on average. 

As counterintuitive as this result might look, they are actually based on a very solid concept in graph theory, namely the emergence of {\it small worlds} from connected networks. A small-world network is a type of mathematical graph in which most nodes are not neighbours of one another, but can be reached from every other one by a small number of steps that actually grows logarithmically with the number of nodes themselves. The six and nineteen degrees of separation highlighted above are different yet similar manifestations of the emergence of small worlds in a network.

Can these concepts be exported to the quantum domain? While the theory of quantum networks has found fertile applications in quantum communication~\cite{internet} and ground-breaking results in the proposal of quantum repeaters for the faithful long-haul transport of quantum information \cite{munro,epping}, the implications of the emergence of small worlds have been far less explored, and mostly confined to studies of excitation-transport and the analysis of the transition from localised to delocalised regimes in spatially extended interacting-particle models~\cite{dima}.

Here, inspired by the analogy between classical network bonds and the correlations set between two elements of a given network of quantum particles, we aim at exploring different aspects. In particular, motivated by the current experimental state-of-the-art in linear optics, which makes available controllable networks of interconnected information carriers, we address the emergence of typical lengths in the (in general multipartite) entanglement established by a random set of unitary gates applied to the elements of a given graph. In particular, we focus on a particular class of operations and networks, i.e. those typically put in place in the procedure for the creation of so-called cluster states, which are resources for measurement-based quantum computing~\cite{Briegel}. 

Such computational paradigm, which has been demonstrated equivalent to any circuital quantum computing protocol, is of fundamental importance in quantum information processing. Linear-optics measurement-based quantum information processing has emerged as a promising avenue for the exploration of controllable quantum protocols. Encoding and entangling qubits in more than one degree of freedom of photons is a promising avenue for the generation of medium-to-large scale photonic cluster states: hyperentanglement-based protocols have so far allowed for the creation of cluster states of up to 6 qubits \cite{Vallone_6qubit}, which have been used to validate fundamental one-way quantum algorithms~\cite{vallone_deutsch, ciampini}. 




In this paper, by randomising the application of the elementary gates needed to engineer a cluster state of a given size, we induce the establishment of small worlds in the underlying network of a given physical system, and address how the spreading of entanglement across the network itself is affected by the degree of stochasticity of such gates. We unveil an interesting hierarchy with which entanglement appears in subnetworks of growing size: only a sufficient degree of determinism allows for the settling of multipartite entanglement within a given cluster lattice, the threshold for $k$-element entanglement depending neatly on the number of elements $k$ itself. 
Moreover, we illustrate a fundamental difference between the phenomenology illustrated in this paper and recently introduced concepts of classical entanglement percolation~\cite{AcinPercol}. 

The remainder of this paper is organised as follows. In Sec.~\ref{theory} we present randomly generated cluster states as the platform for our investigation; in Sec.~\ref{analysis4} we focus our attention to four-qubit cluster states, presenting a rich analysis on the interplay between stochasticity of the gates used to set the network and the settling of bipartite and multipartite entanglement. In Sec.~\ref{sec_Enlarge} we extend our analysis to larger networks.
\section{Theoretical Framework}
\label{theory}

\begin{figure}[t]
\centering
\includegraphics[width=\linewidth]{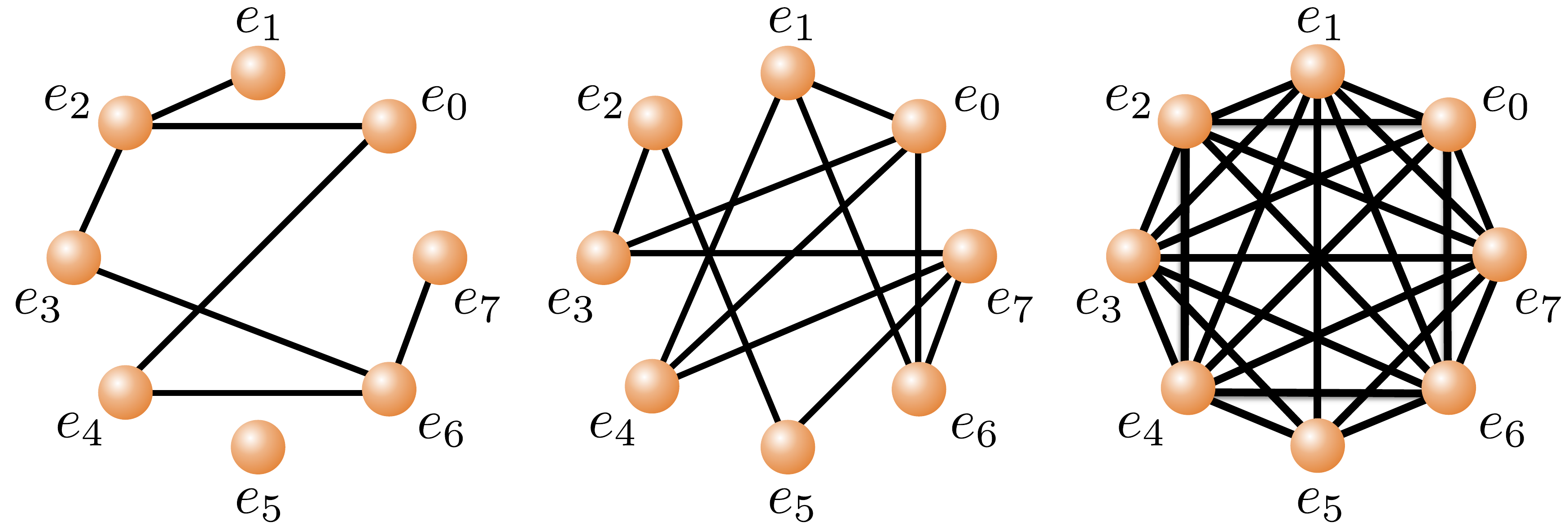}
\caption{Example of instances of an N=8 qubits random cluster states. From leftmost to rightmost panel we have taken $q=0.2$, $q=0.5$, and $q=1$ respectively.}
\label{fig:RandomCluster}
\end{figure}

The approach that we use in order to investigate the core question of our work can be schematised as follows.
\begin{enumerate}
\item We set the value of the threshold $q$ and generate a suitable number of random variables $p_{ij}\in[0,1]$, which embody the probabilities to apply the gate $\cp{i}{j}(\pi)$ to the pair of qubits $(e_i,e_j)$.
\item We compare $p_{ij}$ to $q$. Should it be $p_{ij}<q$ ($p_{ij}>q$), $\cp{i}{j}(\pi)$ is (not) applied. We exhaust the number of all inequivalent pairs of qubits in the network. This produces the network state $\ket{\psi}_{\Sigma}$, where $\Sigma=\{e_1,\dots,e_N\}$ is the set of qubits of the register.
\item We compute the reduced density matrices $\rho_{\sigma}={\rm Tr}_{\Sigma\backslash\sigma}\left[\ket{\psi}\bra{\psi}_\Sigma\right]$ that are obtained upon tracing the overall state over all qubits but those in the subset $\sigma\in\Sigma$.
\item We calculate the percent fraction of such reductions that are entangled at the set value of $q$.
\item In order to eliminate any dependence on the specific random pattern of applications of the joint gate, we repeat the procedure above for a number $Q\gg1$ of instances. 
\item When $Q$ is reached, we change $q$ and repeat the protocol from point $1.$ to $5.$
\end{enumerate}
Needless to say, the number of applications of $\cp{i}{j}(\pi)$ at a set value of the threshold depends strongly on the actual value of $q$ itself: the larger the chosen value of $q$, the higher the number of gate applications. This is illustrated in Fig.~\ref{fig:RandomCluster}, where we show the different configurations achieved for a network of $N=8$ elements for $q=0.2, 0.5$ and $1$, which is associated with a fully connected graph. It is important to remark that, in our notation as well as in Fig.~\ref{fig:RandomCluster}, a bond connecting elements $e_i$ and $e_j$ only means that gate $\cp{i}{j}(\pi)$ was applied, and does not imply the existence of entanglement between such elements. 

Scope of our investigation is ascertaining the phenomenology of distribution of (in general) multipartite entanglement across a given network. In particular, we will focus on the possible emergence of special values of $q$ that are associated with the onset of multipartite entanglement, and the characterisation of such quantum correlations. 

\section{Analysis of the entanglement structure in a random four-qubit state}
\label{analysis4}

We start our analysis by focusing on an intuitive figure of merit that is nevertheless able to provide crucial information on the distribution of entanglement across one of the random graph states discussed above, namely state purity.  We thus proceed to compute the purity
\begin{equation}
{\cal P}_\sigma={\rm Tr}_\sigma[\rho^2_{\sigma}]\in[0,1]
\end{equation}
of the reduced density matrix $\rho_\sigma$, and use the fact that, given the overall pure nature of $\ket{\psi}_\Sigma$, a value of ${\cal P}_{\sigma}<1$ necessarily implies entanglement in the bipartition $(\Sigma\backslash\sigma)\vert\sigma$. We have thus implemented the protocol illustrated in Sec.~\ref{theory} by calculating, in step 4., the percentage of reductions with ${\cal P}_\sigma<1$.

\begin{figure}[b!]
\centering
\includegraphics[width=\columnwidth]{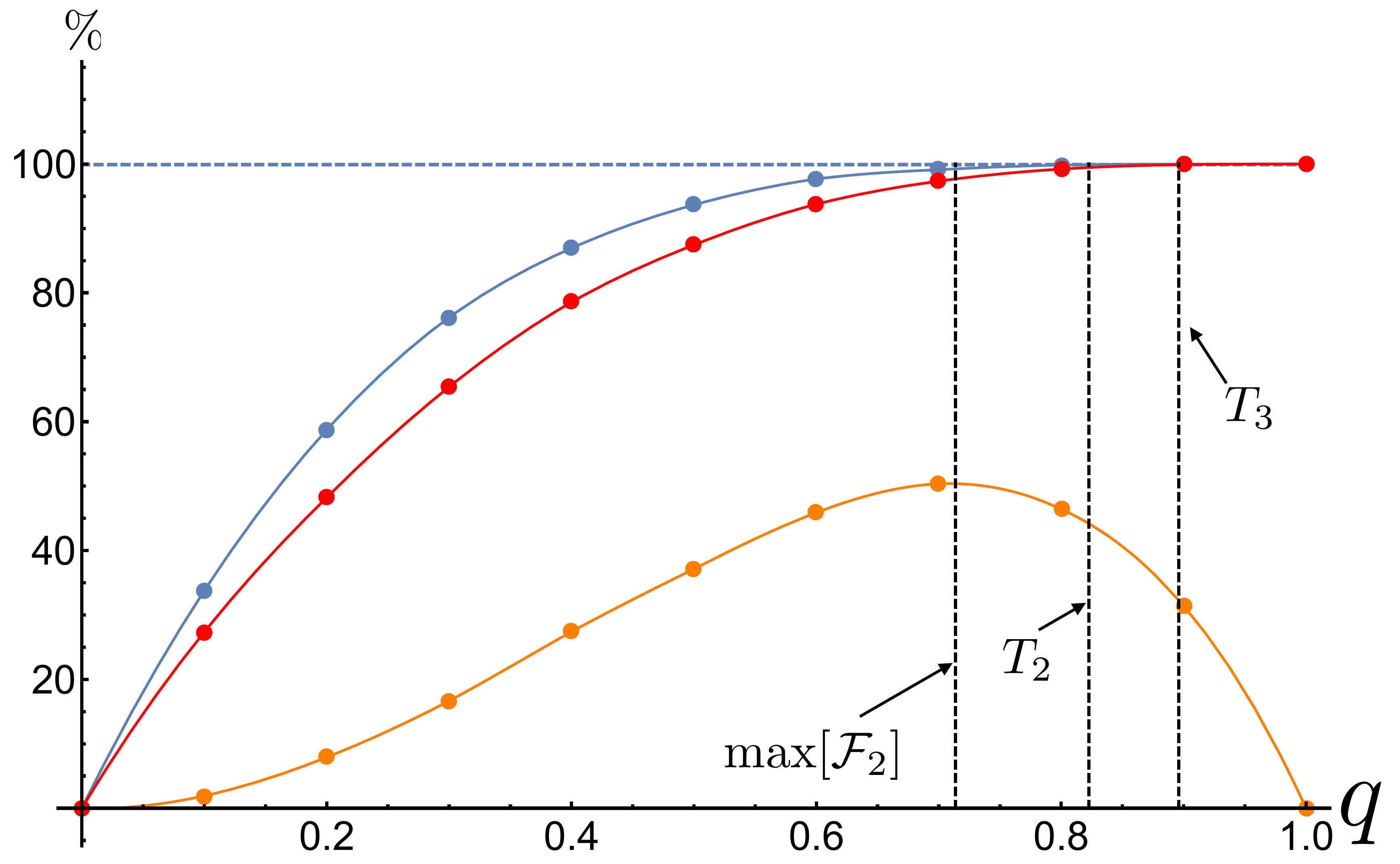}
\caption{We study the percentage fraction of mixed-state reductions that can be identified in a network of $N=4 $ elements, against the threshold parameter $q$. The blue (red) dots show the results of the numerical experiment aimed at quantifying the fraction of mixed two-qubit (one-qubit) reductions. The orange points identify the values of the percentage fraction ${\cal F}_2$ of two-qubit reductions whose purity is exactly $1/4$. The solid lines are non-linear interpolations of the data points.  Each point is the result of an average over a sample of $Q=5000$ elements. Error bars show the standard deviations associated with such averages. Dashed lines $T_{2,3}$ identify the value of $q$ at which the number of mixed two- and one-qubit reductions is at least $99.9\%$ of the possible ones. The line labelled $\max[{\cal F}_2]$ identifies the value of $q$ at which the maximum of ${\cal F}_2$ occurs.}
\label{fig:Mauro4}
\end{figure}

In order to illustrate the salient features of our analysis, we now address explicitly the case of $N=4$, for which $\Sigma=\{e_1,\dots,e_4\}$. The state that would be produced by applying $\cp{i}{j}(\pi)$ gates to every pair of qubits in the network, which would correspond to chosing $q=1$, reads
\begin{equation}
\label{exp}
\begin{aligned}
\ket{\psi}_\Sigma&=\frac{1}{\sqrt2}\hat H_{e_4}(\ket{\phi_+}_{e_1e_4}\ket{\phi_-}_{e_2e_3}+\ket{\psi_+}_{e_1e_4}\ket{\psi_-}_{e_2e_3})\\
&=\frac{1}{\sqrt2}\hat H_{e_3}(\ket{\phi_-}_{e_1e_2}\ket{\phi_+}_{e_3e_4}-\ket{\psi_+}_{e_1e_2}\ket{\psi_-}_{e_3e_4})\\
&=\frac{1}{\sqrt2}\hat H_{e_2}(\ket{\phi_-}_{e_1e_3}\ket{\phi_+}_{e_2e_4}-\ket{\psi_+}_{e_1e_2}\ket{\psi_-}_{e_2e_4})\\
&=\frac{1}{\sqrt2}\hat H_{e_1}(\ket{\phi_-}_{e_1e_2}\ket{\phi_+}_{e_3e_4}-\ket{\psi_+}_{e_1e_2}\ket{\psi_-}_{e_3e_4})
\end{aligned}
\end{equation}
where $\hat H_{e_j}$ is the Hadamard gate on qubit $e_j$ and we have introduced the Bell states $\ket{\phi_\pm}_{e_ie_j}=(\ket{00}\pm\ket{11})_{e_ie_j}/\sqrt2$, $\ket{\psi_\pm}_{e_ie_j}=(\ket{01}\pm\ket{10})_{e_ie_j}/\sqrt2$. The orthogonality of Bell states ensures that entanglement exists in the three inequivalent bipartition $(e_i,e_j)\vert(e_k,e_l)$. Moreover, it is equally straightforward to check that any single-qubit reduction is maximally mixed. Therefore, also the bipartitions $e_i\vert(e_j,e_k,e_l)$ are entangled. This implies that for $q=1$ we expect all six bipartitions that can be identified to be inseparable and the state to be genuinely multipartite entangled. The purity of the associated reduced states is thus necessarily smaller than one. However, for $q<1$ the number of mixed-state reduction is not necessarily as large as six, and our calculations aim at quantifying the percentage of such reduced states as $q$ is varied. 

The results of such calculations are presented in Fig.~\ref{fig:Mauro4} (blue and red dots), where each data point is the result of an average over $Q=5000$ random instances, a sample-size that was large enough to ensure convergence of the numerics. The error bars attached to each point show the uncertainty associated to the averages, calculated as the standard deviation of each $Q$-sized sample and divided by $\sqrt{Q}$. Clearly, for $q=0$ the state of the network is deterministically found to be the factorised initial state $\otimes^4_{j=1}\ket{+}_{e_j}$, while for $q=1$ we retrieve the result anticipated above [cf. Eq.~\eqref{exp}]. In between such extreme situations, the number of inseparable two-vs.-two and one-vs.-three qubits bipartitions (equivalently, mixed two-qubit and one-qubit states) grows monotonically with $q$, albeit at slightly different rates. In particular, we find that the percentage fraction of inseparable two-vs-two (three-vs-one) qubits bipartitions exceeds $99.9\%$ at $q=0.82\pm0.01$ ($q=0.89\pm0.01$), as shown by the vertical dashed line marked as $T_2$ ($T_3$) in Fig.~\ref{fig:Mauro4}. The nominal positions (uncertainties) of $T_{2,3}$ have been obtained as the average (standard deviations) over 100 analytical non-linear interpolations of the results of our simulations, each producing the functions $f_{2,3}(q)$ (whose averages are shown by the blue and red lines in Fig.~\ref{fig:Mauro4}) that have been used to solve numerically the equations $f_{2,3}(q)=99.9$. Quite clearly, $T_2\neq T_3$ beyond statistical errors, which implies that the random network at hand requires a higher threshold in $q$ to produce a complete set of inseparable one-vs-three qubits bipartitions. 

Needless to say, the empirical rule of ``no free lunch" applies here as well: the establishment of multipartite entanglement in the network under scrutiny has to come at the expenses of something else, in light of the monogamy of entanglement. The specific algorithm at hand allows us to explore who pays the toll represented by the establishment of genuine multipartite entanglement in the random network. 

In particular, we expect bipartite entanglement to be affected by the emergence of multipartite one. Such expectation is corroborated by the analysis summarized by the orange dots and curve in Fig.~\ref{fig:Mauro4},  which show the percentage fraction ${\cal F}_2$ of two-vs.-two qubits reductions of random states at a given value of $q$ that have purity exactly equal to $1/4$, which is the lowest a two-qubit state can achieve and witnesses maximum entanglement across the $(e_i,e_j)\vert(e_k,e_l)$ bipartition. Quite intuitively, ${\cal F}_2$ grows at small values of $q$: a low threshold implies very small probability to apply multiple CPHASE gates, which inevitably favours the construction of maximally entangled two-qubit states. For $q\simeq1$, we have a large probability that one qubit is affected by multiple CPHASE gates. Intuitively, this should be able to set strong multipartite entanglement and deplete the degree of bipartite one, and we expect ${\cal F}_2$ to decrease accordingly. Indeed, we know that at $q=1$ we have a genuinely multipartite entangled. The orange dots in Fig.~\ref{fig:Mauro4} confirm such expectation, and show the occurrence of a maximum of ${\cal F}_2$ that is close, yet not identical, to the chosen thresholds $T_{2,3}$ discussed above (we have that $\max[{\cal F}_2]$ occurs at $q=0.72\pm0.01$). 

\begin{figure}[t]
\centering
\includegraphics[width=\linewidth]{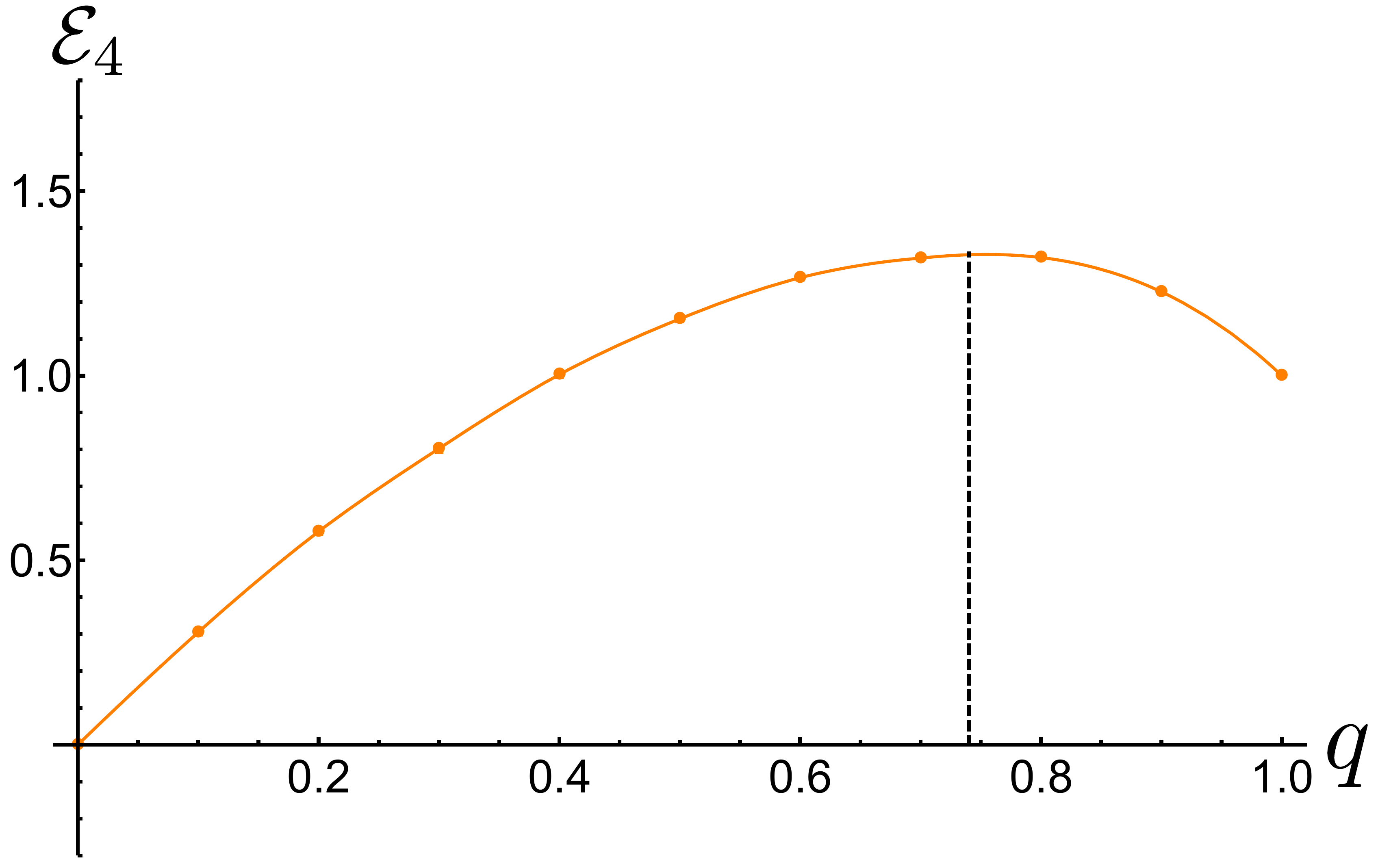}
\caption{Average four-partite negativity ${\cal E}_4$ plotted against $q$ obtained for a sample of $Q=5000$ random network states. The error bars are the standard deviations associated with the averages. The orange solid line is a non-linear interpolating function whose maximum is achieved at $q=0.72\pm0.01$ (vertical dashed line).}
\label{Entanglement4}
\end{figure}

Of course, counting for the number of reductions that are in mixed states does not provide full information about the actual degree of multipartite entanglement that is established among the elements of the network. Only the assessment of a figure of merit for genuine multipartite entanglement can give us insight into. We have thus decided to use the $N-$partite generalisation of negativity as a quantifier of the degree of entanglement within our resource state. This is defined as
\begin{equation}
\label{NpartiteNeg}
{\cal E}_N=\sqrt[N]{\Pi_{\{\sigma\}}{\cal E}_{\sigma\vert\Sigma\backslash\sigma}}
\end{equation} 
where ${\cal E}_{\sigma\vert\Sigma\backslash\sigma}$ is the negativity of the partially transposed density matrix of the bipartition $\sigma\vert\Sigma\backslash\sigma$ and the product extends to all the bipartitions. We recall the definition of negativity as 
\begin{equation}
{\cal E}_{\sigma\vert\Sigma\backslash\sigma}=\max[0,-2\sum_j\lambda^-_j]
\end{equation}
with $\{\lambda^-_j\}$ the set of negative eigenvalues of the partially transposed (with respect to any of the subparties) density matrix of the bipartition $\sigma\vert\Sigma\backslash\sigma$. The geometric average upon which Eq.~\eqref{NpartiteNeg} is built is null whenever at least one of the bipartitions of the network is positive under partial transposition. Therefore, only if all bipartitions are certified inseparable according to the partial transposition criterion is the state of the network genuinely multipartite entangled. While this does not exclude the possibility of facing bound entanglement, the problem addressed in this work never presents states that are positive under partial transposition (with the exception of the state achieved for $q=0$, which is identical to the factorised initial state). Fig.~\ref{Entanglement4} shows the behavior of ${\cal E}_4$ against $q$. While for $q>0$ we always have four-partite entanglement (in line with the finding in Fig.~\ref{fig:Mauro4}), it is remarkable that $q=1$ is not associated with the largest degree of four-partite negativity, which actually occurs at $q=0.72\pm0.01$. 

\begin{figure}[t]
\centering
\includegraphics[width=\linewidth]{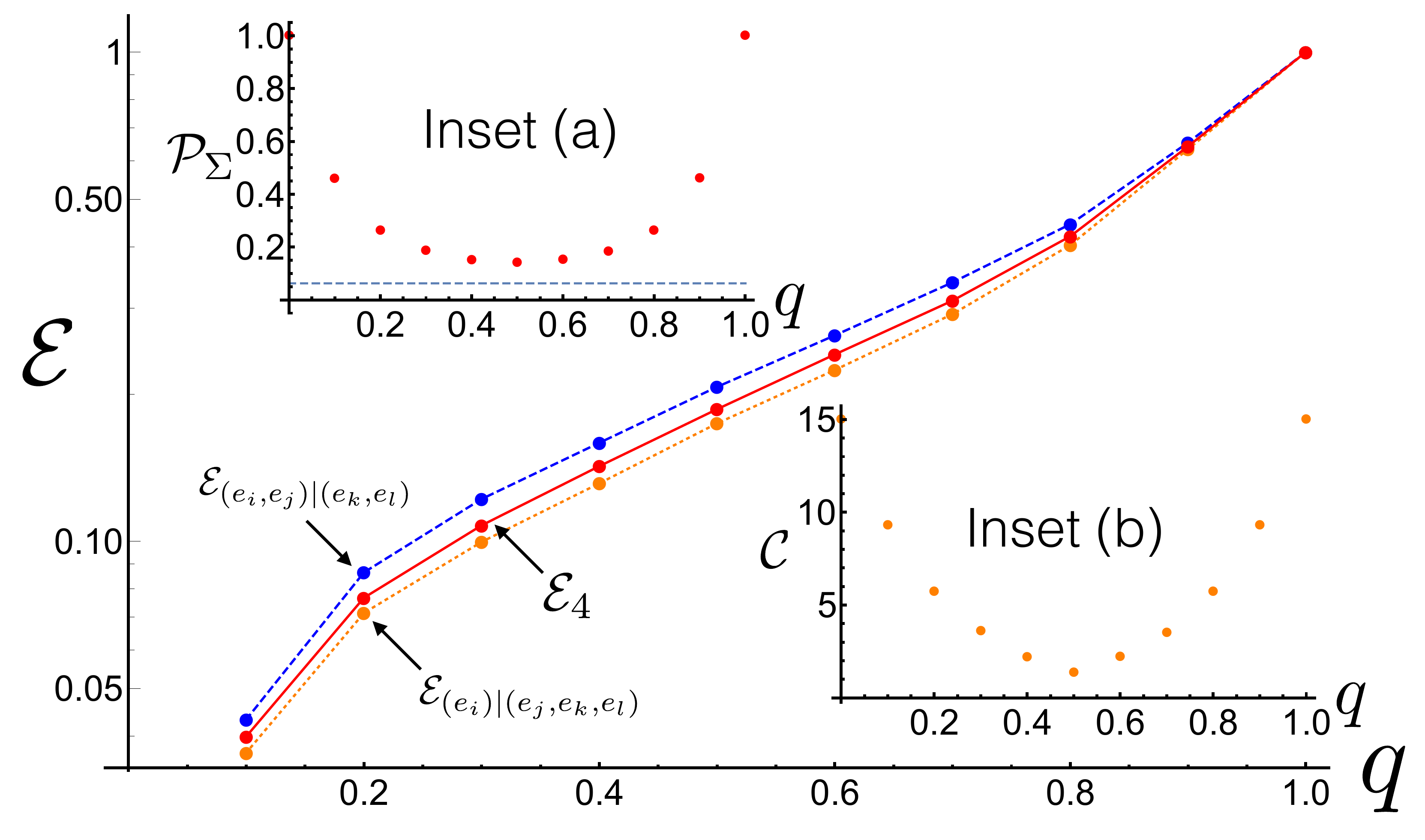}
\caption{{\bf Main panel}: Logarithmic plot of the entanglement within the average estate $\rho_\Sigma$ of an $N=4$ random network against the threshold probability $q$. The red dots show the value taken by the four-partite negativity ${\cal E}_4$, while the blue and orange ones are for the entanglement within the bipartitions $(e_i,e_j)\vert(e_k,e_l)$ and $(e_i)\vert(e_j,e_k,e_l)$. The lines connecting the dots are simply guides to the eye. {\bf Inset (a)}: Purity ${\cal P}_\Sigma$ of the average state against $q$. The dashed horizontal line shows the minimum purity of a four-qubit state. {\bf Inset (b)}: Values taken by the measure of coherence ${\cal C}$ against the threshold probability.}
\label{StatoMedio4}
\end{figure}

We continue the assessment of the four-partite case by pointing out the differences between the average behavior of the figures of merit addressed herein and the values taken by such indicators over the {\it average state} of the network. The latter is defined as the state obtained upon mediating over $Q$ random instances of network states. Formally, by assuming all instances to be equally likely to occur (which is entailed by choosing the probabilities to apply gates $\cp{i}{j}(\pi)$ uniformly), the physical state of the system is described by the density matrix
\begin{equation}
\label{medio}
\rho_\Sigma=\frac{1}{Q}\sum^Q_{j=1}\ket{\psi}\bra{\psi}_{\Sigma,j},
\end{equation}
where $\ket{\psi}\bra{\psi}_{\Sigma,j}$ is the $j^\text{th}$ random state of the $Q$-sized sample.
With the exception of the cases associated with $q=0,1$ (when we sum identically prepared states), by averaging we lose the purity of the network state: ${\cal P}_\Sigma$ reaches values as low as $\simeq0.14$ for $q=0.5$ [cf. Inset (a) of Fig.~\ref{StatoMedio4}], which is however larger than the minimum purity $1/16$ achievable by a four-qubit state. Despite being mixed, the average state of the network preserves significant quantum coherences as quantified by the measure proposed in Ref.~\cite{Coherence} and fomalised as
\begin{equation}
{\cal C}=\sum_{i\neq j}\vert(\rho_\Sigma)_{ij}\vert
\end{equation}
with $\vert(\rho_\Sigma)_{ij}\vert$ the off-diagonal elements of the density matrix $\rho_\Sigma$. The behavior of ${\cal C}$ against $q$ is shown in Inset (b) in Fig.~\ref{StatoMedio4}: a minimum of the measure of coherence is achieved in correspondence of the minimum purity. However, such a minimum is strictly non-null, thus leaving open the possibility of dealing with a (mixed) entangled state of the network. Such a possibility is confirmed by the analysis of the four-partite negativity ${\cal E}_4$ [cf. main panel of Fig.~\ref{StatoMedio4}], which is a growing function of $q$ [similar trends are exhibited by both the two-vs.-two qubits entanglement ${\cal E}_{(e_i,e_j)\vert(e_k,e_l)}$, and the one-vs.three qubits one ${\cal E}_{(e_i)\vert(e_j,e_k,e_l)}$]. Nothing remarkable in the behavior of ${\cal E}_4$ appears to be related to the value of $q=0.5$, although the entanglement function changes concavity in correspondence to such a value of the probability threshold. 
  
\begin{figure}[t]
\centering
\includegraphics[width=\linewidth]{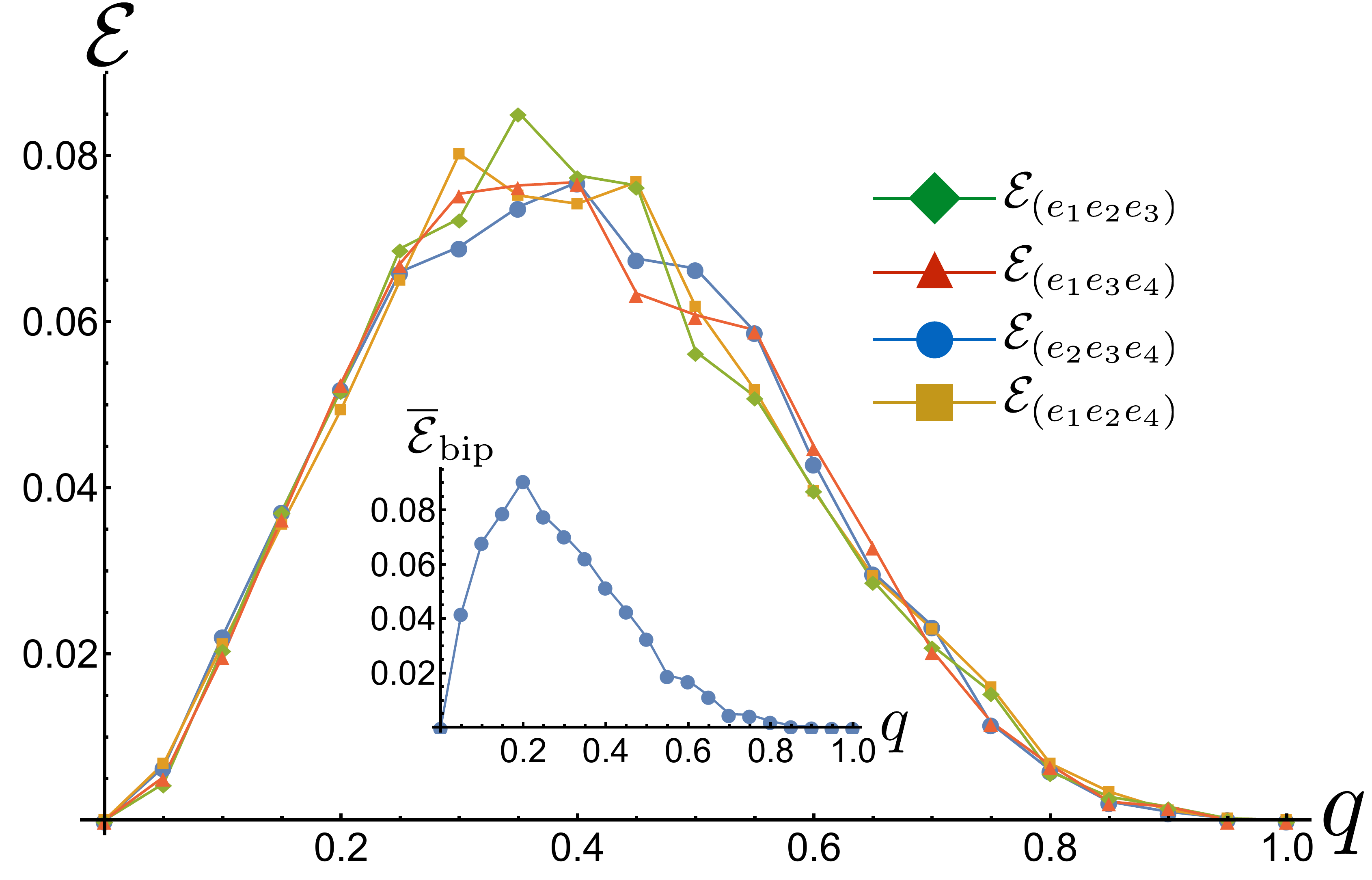}
\caption{{\bf Main panel}: Tripartite negativity in the three-qubit reductions identified in the legend, plotted against $q$. Each plot is an average over $Q=5000$ realisation of the random network state (we omit the error bars for clarity of presentation).  {\bf Inset}: Mean bipartite negativity $\overline{\cal E}_{\text{bip}}$ averaged over the 6 two-qubit reduced states that can be singled out from our network. Same conditions as in the main panel.}
\label{Reductions4}
\end{figure}

To finish the study of this paradigmatic case, we report in the main panel of Fig.~\ref{Reductions4} the behavior of the tripartite entanglement in the four three-qubit reduced states that can be singled out from our network. We have used the tripartite version of Eq.~\eqref{NpartiteNeg} to quantify the entanglement and changed our notation so as to make explicit the triplets of elements of the network that we ave considered. Moreover, by tracing out two elements, we have evaluated the residual two-qubit entanglement, whose average across the six two-qubit reductions is displayed in the inset of Fig.~\ref{Reductions4}. The general trend of such figures of merit follows the expectation that, in the large-$q$ region, the entanglement in the reduction is depleted to favour the emergence of genuinely multipartite one. Moreover, their quantitative value is, in general, very small. A point of notice is that the peak of three- and two-qubit negativity does not occur at the same value of $q$, thus suggesting an interesting hierarchy of values of $q$ at which the various structures of entanglement across the system are triggered or destroyed.

\begin{figure}[t]
\centering
\includegraphics[width=\linewidth]{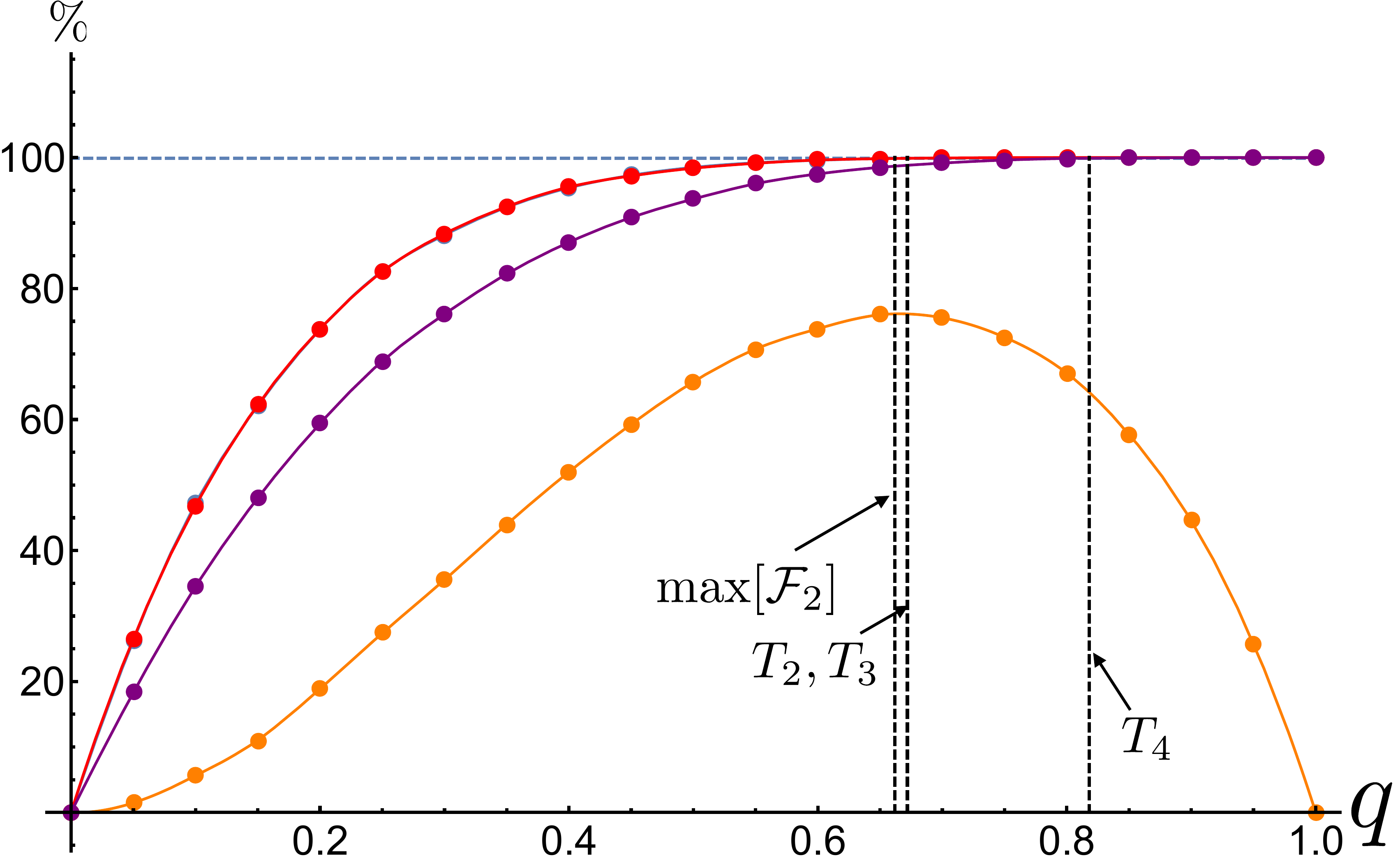}
\caption{We study the percentage fraction of mixed-state reductions that can be identified in a network of $N=5$ elements, against the threshold parameter $q$. The red dots show the results of the numerical experiment aimed at quantifying the fraction of mixed two- and three-qubit reductions, which actually coincide. The purple dots show the results for the one-qubit reductions.The orange points identify the values of the percentage fraction ${\cal F}_2$ of two-qubit reductions whose purity is exactly $1/4$. The solid lines are non-linear interpolations of the data points.  Each point is the result of an average over a sample of $Q=10^4$ elements. Error bars show the standard deviations associated with such averages. Dashed lines $T_{2,3}$ ($T_4)$ identify the value of $q$ at which the number of mixed two- and three-qubit (one-qubit) reductions is at least $99.9\%$ of the possible ones. The line labelled $\max[{\cal F}_2]$ identifies the value of $q$ at which the maximum of ${\cal F}_2$ occurs.}
\label{Mauro5}
\end{figure}

\section{Enlarging the size of the network}
\label{sec_Enlarge}
\begin{figure}[t]
\centering
\includegraphics[width=\linewidth]{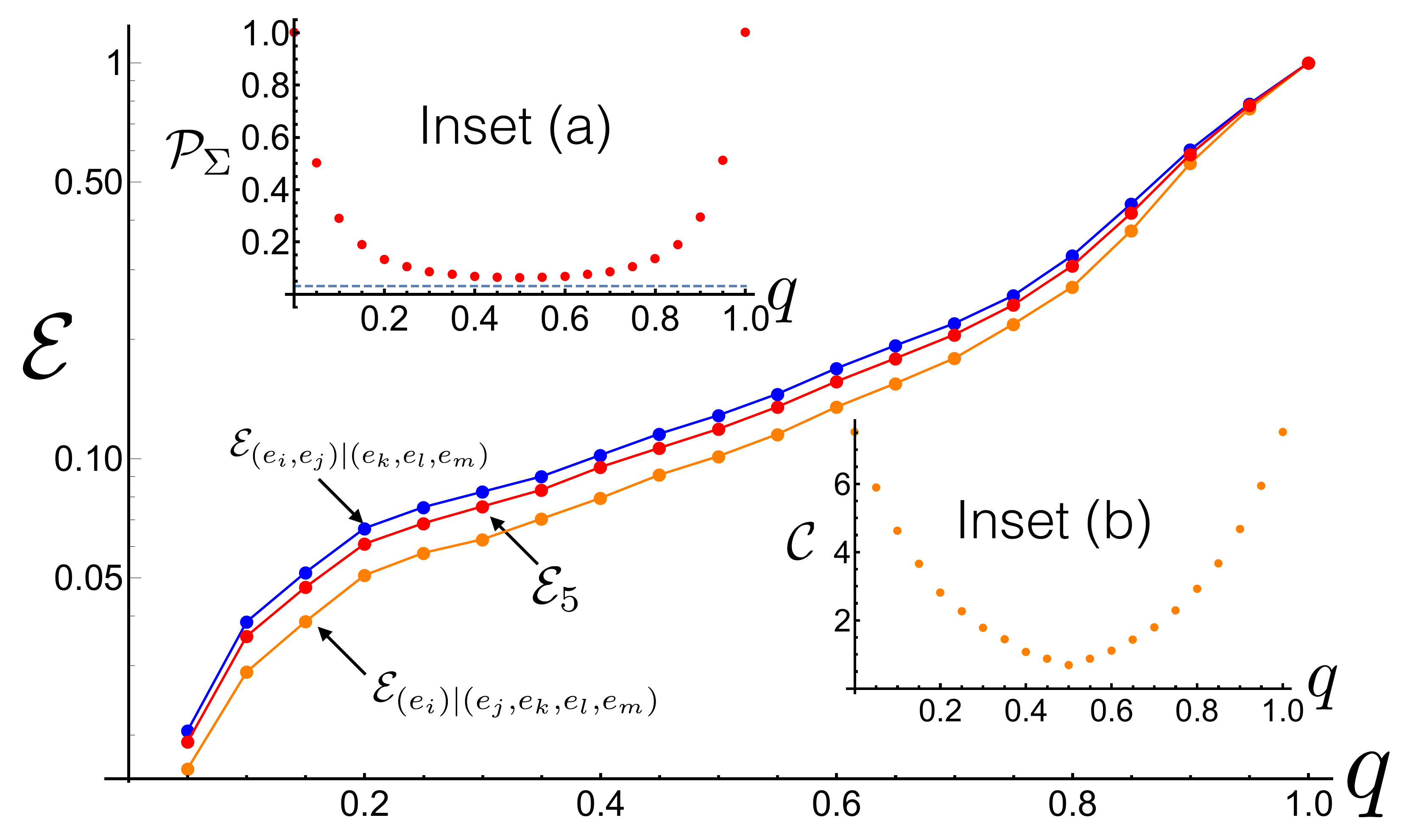}
\caption{{\bf Main panel}: Logarithmic plot of the entanglement within the average estate $\rho_\Sigma$ of an $N=5$ random network against the threshold probability $q$. The red dots show the value taken by the four-partite negativity ${\cal E}_5$, while the blue and orange ones are for the entanglement within the bipartitions $(e_i,e_j)\vert(e_k,e_l,e_m)$ and $(e_i)\vert(e_j,e_k,e_l,e_m)$. The lines connecting the dots are simply guides to the eye. {\bf Inset (a)}: Purity ${\cal P}_\Sigma$ of the average state against $q$. The dashed horizontal line shows the minimum purity of a four-qubit state. {\bf Inset (b)}: Values taken by the measure of coherence ${\cal C}$ against the threshold probability.}
\label{StatoMedio5}
\end{figure}

We now assess the features of larger networks of qubits, addressing questions that are akin to those assessed in Sec.~\ref{analysis4}. Features similar to those showcased in the four-qubit network are present in all the higher-dimensional systems that we have studied through our simulations. For instance, Figs.~\ref{Mauro5} and \ref{StatoMedio5} display the same behaviors highlighted in Figs.~\ref{fig:Mauro4} and \ref{StatoMedio4}, respectively. Rather than reporting qualitatively similar plots for larger networks, in Table~\ref{tavola} we present the threshold values of $q$ at which progressively larger reductions of the state of the network are mixed. 

The trend is clear: as we look into larger networks, the value of $T_k~(k=2,3,\dots)$ decreases. 
\begin{table}[b]
\begin{tabular}{ccccccc}
\hline\hline
${N}$&$4$&$5$&$6$&$\cdots$&$9$\\ 
\hline\hline
$\max{\cal F}_2$&0.72&0.66&0.64&&0.40\\
$T_2$&$0.82$&$0.67$&$0.57$&&0.39\\
$T_3$&$0.89$&$0.67$&$0.54$&&0.31\\
$T_4$&$\blacksquare$&0.818&0.57&&0.27\\
$T_5$&$\blacksquare$&$\blacksquare$&0.75&&0.27\\
$T_6$&$\blacksquare$&$\blacksquare$&$\blacksquare$&&0.31\\
$T_7$&$\blacksquare$&$\blacksquare$&$\blacksquare$&&0.39\\
$T_8$&$\blacksquare$&$\blacksquare$&$\blacksquare$&&0.40\\

 \hline\hline
\end{tabular}
\caption{\label{tavola}The table shows the threshold value of $q$ at which the fraction of progressively larger reductions in an $N$-element random network is at least $99.9\%$. Black squares stands for unavailable data at that size of the network. As before, $\max[{\cal F}_2]$ is the value of $q$ at which the maximum of ${\cal F}_2$ occurs.}
\end{table}

\subsection{Entanglement Percolation}

It is interesting to compare our analysis to  entanglement percolation, a concept akin to classical bond percolation introduced in Ref.~\cite{AcinPercol}. Consider a graph of particles akin to one of those addressed in this paper. This time, though, a link between two elements implies the presence of entanglement between them. Refs.~\cite{AcinPercol} show the existence of a 
minimum amount of entanglement between any two elements of the network needed to establish a perfect  quantum channel between distant (not directly connected) elements, with significant (non-exponentially decaying) probability.

This is fundamentally different from our situation, where instead we point out the existence of a minimum probability to randomly apply a two-qubit gate in a network associated with the establishment of a genuinely multipartite entangled state of the network. Our threshold does not guarantee the existence of a long-distance entangled channel between arbitrarily chosen elements of the network. In fact, non-nearest-neighbour elements of a cluster state are not necessarily entangled, their entanglement being in general dependent on the geometry of the underlying network. 

\begin{figure}[t]
\centering
\includegraphics[width=0.98\linewidth]{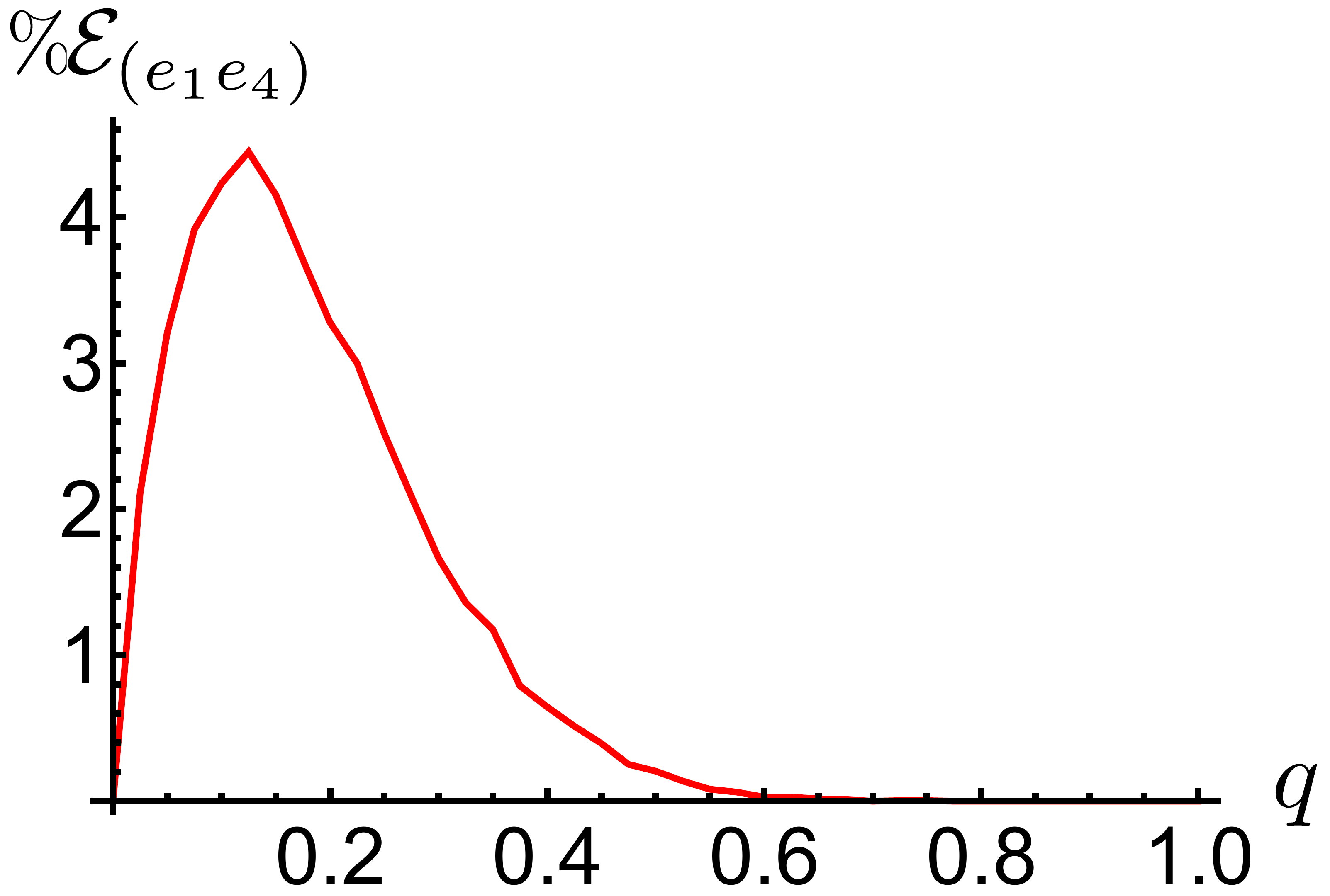}
\caption{Percentage of reduced states of elements $e_1$ and $e_4$ of an $N=6$ random network exhibiting a non-zero negativity, plotted against $q$, for a sample of 5$\times10^4$ states. }
\label{checkperco}
\end{figure}

In order to ascertain if a value of $q$ exists above which long-haul entanglement is set in the network, we computed the negativity of the reduced state of the qubits that have the largest number of intermediate sites between them, at a given value of $N$. This is analogous to the study presented in the inset of Fig.~\ref{Reductions4}, although instead of an average over all the possible two-qubit reductions, here we consider now only a specific reduction. Fig.~\ref{checkperco} shows the results valid for the case of $N=6$, for which we address the entanglement between elements $e_1$ and $e_4$. We have considered the percentage of reductions of such elements with a non-zero value of negativity against the value of $q$. Quite clearly, such a percentage remains always very small, regardless of $q$, showing that no classical entanglement percolation effect occurs, as there is no value of $q$ at which long-distance entanglement within the network is set deterministically.  The results should be considered as canonical, qualitatively valid regardless of the actual choice of $N$, and indicative of the profound differences between the situation addressed here and the study in Ref.~\cite{AcinPercol}. 

\section{\label{sec:four}Conclusions}
We have studied the entanglement sharing structure among the elements of a qubit network subjected to probabilistic CPHASE gates. We have highlighted the existence of statistically inequivalent thresholds in the probability of application of the gates for the settling of entanglement in various subsets of network elements, thus unveiling an interesting hierarchy in the entanglement distribution pattern of a given network. The phenomenology that we have highlighted cannot be understood in terms of the statistical properties of an intuitive, yet too naive, reference state such as the one obtained by averaging overall the elements of the random set of states generated in our numerical experiments: the above-mentioned hierarchy is a statistical feature of random networks rather than a property of the statistically average state of the network. Remarkably, {\it small worlds} structures in the entanglement sharing of the random set of network states appear to emerge. This is an interesting feature that deserves more attention and upon which we plan to focus our forthcoming (theoretical and experimental) efforts.



\begin{thebibliography}{99}

\bibitem{6} F. Karinthy, {\it L\'ancszemek}, in {\it Minden mask\'eppen van} (1929).

\bibitem{barabba} R. Albert, H. jeong, and A>-L. Barabasi, Nature {\bf 401}, 130 (1999).

\bibitem{internet} H. J. Kimble, Nature {\bf 453}, 1023 (2008). 

\bibitem{munro} W. J. Munro, K. A. Harrison, A. M. Stephens, S. J. Devitt, and K. Nemoto, Nat. Photonics {\bf 4}, 792-796 (2010).

\bibitem{epping} M. Epping, H. Kampermann, and D. Bru{\ss}, New J. Phys. {\bf 18}, (2016)

\bibitem{dima} C. P. Zhu and S.-J. Xiong, Phys. Rev. B {\bf 62}, 14780 (2000); O. Giraud, B. Georgeot, and D. L. Shepelyansky, Phys. Rev. E {\bf 72}, 036203 (2005).

\bibitem{Briegel} H. J. Briegel, D. E. Browne, W. D\"ur, R. Raussendorf, M. Van den Nest, Nature Phys. {\bf 5}, 19 (2009).




\bibitem{Vallone_6qubit}{G. Vallone, G. Donati, R. Ceccarelli,  and P. Mataloni, Phys. Rev. A, {\bf 81} 052301 (2010)}

\bibitem{vallone_deutsch}{G. Vallone, E. Pomarico, F. De Martini, and P. Mataloni, Phys. Rev. A {\bf 78}, 042335 (2008).}


\bibitem{ciampini}{M. A. Ciampini, A. Orieux, S. Paesani, F. Sciarrino, G. Corrielli, A. Crespi, R. Ramponi, R. Osellame, P. Mataloni, Light Sci. Appl, {\bf 5}, e16064 (2016).
}


\bibitem{AcinPercol} A. Ac\'in, J. I. Cirac, and M. Lewenstein, Nature Phys. {\bf 3}, 256 (2007). 




\bibitem{Coherence} T. Baumgratz, M. Cramer, and M. B. Plenio, Phys. Rev. Lett. {\bf 113}, 140401 (2014).


\end{thebibliography}
\end{document}